\documentclass[aps,pre,superscriptaddress,letterpaper,twocolumn,longbibliography]{revtex4-2}

\usepackage[utf8]{inputenc}
\usepackage{mwe}
\usepackage{graphicx}% Include figure files
\usepackage{dcolumn}% Align table columns on decimal point
\usepackage{bm}% bold math
\usepackage{bbm}
\usepackage{hyperref}% add hypertext capabilities

\usepackage{xcolor}
\usepackage{amsmath}
\usepackage{amssymb}
\usepackage{xcolor}
\usepackage{svg}

\begin{document}

\title{Exploring Spatial Segregation Induced by Competition Avoidance as Driving Mechanism for Emergent Coexistence in Microbial Communities}

\author{Mattia Mattei}
\affiliation{\small{Departament d'Enginyeria Inform{\`a}tica i Matem{\`a}tiques,Universitat Rovira i Virgili, 43007 Tarragona, Spain}}
\author{Alex Arenas}
\affiliation{\small{Departament d'Enginyeria Inform{\`a}tica i Matem{\`a}tiques,Universitat Rovira i Virgili, 43007 Tarragona, Spain}}
\affiliation
{Pacific Northwest National Laboratory, 902 Battelle Blvd, Richland, WA, 99354, USA}

\begin{abstract}
This study investigates the role of spatial segregation, prompted by competition avoidance, as a key mechanism for emergent coexistence within microbial communities. Recognizing these communities as complex adaptive systems, we challenge the sufficiency of mean-field pairwise interaction models and consider the impact of spatial dynamics. We developed an individual-based spatial simulation depicting bacterial movement through a pattern of random walks influenced by competition avoidance, leading to the formation of spatially segregated clusters. This model was integrated with a Lotka-Volterra metapopulation framework focused on competitive interactions. Our findings reveal that spatial segregation combined with low diffusion rates and high compositional heterogeneity among patches can lead to emergent coexistence in microbial communities. This reveals a novel mechanism underpinning the formation of stable, coexisting microbe clusters, which is nonetheless incapable of promoting coexistence in the case of isolated pairs of species. This study underscores the importance of considering spatial factors in understanding the dynamics of microbial ecosystems.  
\end{abstract}

%\date{\today}
\maketitle

\section{Introduction}
Microorganisms do not function in isolation; rather, they exist within multi-species communities, cohabiting the same environment and engaging in a spectrum of complex interactions. This complexity poses a substantial challenge in understanding how the physiological behaviors of individual species contribute to emergent properties such as stability, productivity, and resilience. From this perspective, microbial communities can be regarded as `\emph{complex adaptive systems}' \cite{pr2040711}, making them well-suited for rigorous quantitative theoretical analysis. Indeed, a plethora of mathematical models have been applied to describe the microbiome. These range from variations of the Lotka-Volterra equations \cite{doi:10.1126/science.aad2602} and MacArthur's consumer-resource model \cite{doi:10.1126/science.aat1168, Posfai} to frameworks based on evolutionary game theory \cite{FREY20104265}, among others.

Recently, Chang et al. \cite{doi:10.1126/science.adg0727} presented solid experimental proofs that multispecies coexistence is an \emph{emergent phenomenon}: they isolated organisms from stable synthetic bacterial communities consisting of various species and competed all possible combination of pairs of organisms to test their ability to live together; in most cases, one species outcompeted the other, leading to exclusion. From this, they concluded that coexistence in communities cannot be reduced to pairwise coexistence rules and they left open the question about 
the fundamental mechanisms behind it. In particular, they wonder whether the complex nature of multispecies coexistence derives from higher-order interactions (HOIs), or whether it can be explained by a complex network of pairwise relations.

Providing a precise definition of HOIs, particularly in the field of ecology, is challenging \cite{SANCHEZ2019519}. One possible simplified conceptualization is to view HOIs as a modifications of pairwise interactions in the presence of a third or more species. The research by Mickalide and Kuehn \cite{MICKALIDE2019521} reveals a HOI aligning with this definition. An `interaction modification' unfolds, wherein a single-celled algae (Chlamydomonas reinhardtii) alters the dynamics between a predatory ciliate (Tetrahymena thermophila) and the bacterium E. coli. This modification stems from a phenotypic change in E. coli triggered by the presence of C. reinhardtii; the algae inhibits the aggregation of E. coli cells, rendering them more susceptible to predation by the ciliate.

In general, it is always possible to generalize mathematical models to incorporate groups interactions \cite{Grilli}, but identifying the various and often intricate driving physical mechanisms behind them can be demanding. Before employing such sophisticated tools and concepts, we wonder whether the simpler framework of complex networks of pairwise interactions is sufficient to explain the emergent phenomena.

Studies such as those conducted by Thebault and Fontaine \cite{thebault} or Rohr et al. \cite{Rohr} have illustrated how the architecture of the network of pairwise interactions among species significantly influences the stability and overall macroscopic characteristics of an ecosystem. In this context, a pertinent question emerges: why, within the same ecosystem, do certain species interact among them while others do not? Perhaps the most straightforward explanation lies in how species are \emph{spatially distributed}.

There is abundant evidence indicating the significance of spatial constraints in the formation of microbial communities and the emergence of spatially segregated clusters of bacteria. For instance, Welch et al. \cite{Welch} discovered highly organized spatial structure in the oral microbiome and a surprising correlation between position in space of taxa and their function. Conwill et al. \cite{Conwill} showed how lineages with in vitro fitness differences coexist within centimeter-scale regions on human skin, but each skin pore being dominated by a single lineage. Shi et al. \cite{Shi} developed a new technology for mapping the microbiome and they discovered that microbial communities in oral biofilms are spatially structured as stable microarchitectures over time. Cho et al. \cite{Cho}, through the utilization of an innovative microfluidic device, showed that bacterial colonies of \emph{E. Coli} have the capability to autonomously self-organize within chambers of varying shapes and sizes that permit continuous cell escape. In general, colonies of bacteria exhibit a fascinating propensity to form tight structures in various settings
in response to unfavorable environmental conditions including various types of chemical stress \cite{Stoodley}.

The core concept of this study is to explore the potential of spatial self-organization among microbes as the underlying factor contributing to the observed emergent coexistence identified by Chang et al. To achieve this objective, it is necessary to understand bacterial movement. 

 Bacteria in a liquid medium exhibit a movement pattern characterized by alternating between tumble and swim phases \cite{Berg}. In a uniform environment, the movement of a bacterium resembles a random walk, with relatively straight swimming segments occasionally interrupted by random tumbles that reorient the bacterium. Notably, bacteria such as E. coli lack the ability to deliberately choose their swimming direction and cannot maintain a straight path for an extended period due to rotational diffusion, essentially ``forgetting" their trajectory. To compensate for this, they continuously assess their course and make adjustments when necessary, allowing them to steer their random walk towards favorable locations. This movement of bacteria in response to chemical gradient is called \emph{chemotaxis}. The combination of these forces can result in complex colony formation and various types of collective motion, as proved by numerous experiments and theoretical models \cite{Czirok, Alber, Wu, Peruani}.

While it may be reasonable to attribute bacterial movements primarily to nutrient-driven research, it is essential to acknowledge that chemotaxis is both imprecise and energetically costly for bacteria, especially in densely populated environments (Brumley \cite{Brumley}). Additionally, when considering bacterial communities in batch cultures in vitro, such as those studied by Chang et al., there is no clear reason to assume uneven nutrient distribution or strong nutrient gradients influencing bacterial motion. A more plausible proposition would be that the predominant driving force behind bacterial movement is an unoriented escape response from highly competitive environments to less competitive ones, facilitated by a uniform distribution of nutrients.

Following this direction, we developed an individual-based spatial simulation to depict the individual movement of bacteria, leading to the formation of spatially segregated clusters resulting from the \emph{escape} from  regions with high competition. Under the assumption of time-scales separation between movement and growth, the result of the spatial simulation was then utilized to calculate the initial conditions for a metapopulations Lotka-Volterra model with only competitive interactions, capturing the growth dynamics of such patches of bacteria. This study shows that: i) segregation of clusters of bacteria can be obtained as a result of competition avoidance only and, therefore, it can potentially occur in any conditions regardless of the environmental setup; ii) building upon the latter justification, a metapopulation Lotka-Volterra formalism can be adopted and this alters considerably the pattern of coexistence for the species at the equilibrium. Specifically, for low diffusion rates and high compositional heterogeneity among patches, it is possible to reproduce the emergence observed by Chang et al., i.e. the formation of stable communities formed by multiple microbial species, the majority of which do not coexist when isolated in pairwise combinations. As a benchmark to validate our model, we also confirmed its ability to replicate the three macroecological laws governing microbial communities, as discovered by J. Grilli\cite{Grilli1}.

\section{Individual-based Spatial Simulation for Bacterial Motion}

The simulation starts by uniformly distributing $n$ bacteria within a two-dimensional square and subsequently randomly assigning each of them to one of $N<n$ different species. While other initial spatial distributions are conceivable—such as placing bacteria of the same species in closer proximity—these variations do not appear to influence the final results. Hence, we opted to proceed with the initially uniform distribution for its generality. We can then assume that each cell interacts only with the cells in its proximity. If we think about bacteria as nodes in a network, the simple formalism of Random Geometric Graph (RGG) \cite{penrose} can be employed to rapidly evaluate the number of bacteria in the neighborhood of each node. In a RGG two nodes are connected when their distance is within a certain neighborhood radius $R$. Here, the euclidean distance is considered, i.e. two nodes $i$ and $j$ are connected when $d_{ij} = \sqrt{(x_i - x_j)^2 + (y_i - y_j)^2} < R$. The idea is that the higher the density of competitors in the vicinity of node $i$ the more it is inclined to escape from that region. Therefore, we update nodes positions according to:
\begin{equation}
    \begin{cases}
    x_i(t+1) = x_i(t) + I_i(t)\cos{\theta_i(t)},\\
    y_i(t+1) = y_i(t) + I_i(t)\sin{\theta_i(t)},
    \end{cases}
\end{equation}
with $\theta_i$ uniformly distributed between $[0, 2\pi]$ at each time-step and
\begin{equation}
    I_i(t) = \frac{R}{1+e^{-\alpha \Big(\frac{N_c^i(t)}{N_{th}} -1\Big)}}.
    \label{eqn:intensity}
\end{equation}
Stated differently, we model bacterial movement selecting a random direction and an intensity proportional to the level of competitiveness, i.e. the number of competitors $N_c^i$ in the neighborhood of the node $i$. Taking inspiration from neural networks, the intensity of motion is modelled as a sigmoid function. The parameter $\alpha$ controls the the shape of the curve; if $\alpha\to\infty$ the function tends to the Heaviside step-function. The threshold parameter $N_{th}$ controls the position of the inflection point $N_c^i = N_{th}$. This function cannot be equal to zero, disallowing bacteria to be completely still, and its maximum value is equal to $R$, i.e. the radius of the neighborhood area.
%\begin{figure}
%    \centering
%    \includegraphics[scale = 0.55]{Figure1.pdf}
  %  \caption{Bacterial intensity of motion as function of the density of local competitors. Each curve is described by equation \ref{eqn:intensity} with different $\alpha$ between 1 and 20. The greater $\alpha$, the more the function tends to a step function. Here, $R$ is chosen equal to 0.2.}
  %  \label{fig:intensity}
%\end{figure}
The simulation stops when all the bacteria minimize their intensity motion, i.e. when their activity results in a random walk confined in a small portion of space. In order to quantitatively compute the different clusters of bacteria in the final network we employed the classical Louvain Algorithm for communities detection \cite{Blondel_2008}. It optimizes a quality function known as modularity, which quantifies the strength of the community structure in a network. The algorithm optimizes modularity by iteratively moving nodes between communities, enhancing the overall cohesion within communities while reducing connectivity between them. The Louvain algorithm is a popular choice in the field of network science, despite the numerous limitations associated with modularity optimization algorithms \cite{Peixoto}. In our context, the selection of a particular algorithm has negligible effects.

We assume that the spatial rearrangement occurs rapidly enough to achieve equilibrium before species start to grow. This separation of temporal scales allows us to model growth using a classical population-based ecological model as the Lotka-Volterra formalism, adapted to our context as described in the following section.

\section{Metapopulation Lotka-Volterra Model}

The Lotka-Volterra equations are the gold standard to model the dynamics of interacting populations in ecology \cite{Volterra}. In the generalized version for $N$ species, they can reproduce each possible type of relations according to the sign of the interactions matrix entries, from competition to cooperation. A growing debate surrounds the significance of positive interactions among bacterial species. Numerous studies, as the one by Palmer and Foster \cite{Palmer}, showed that negative interactions tend to predominate, and instances of cooperation, where two species mutually benefit, are generally infrequent. Contrary findings are presented in studies such as that of Kehe et al. \cite{kehe}, wherein positive interactions, particularly parasitisms, are identified as common occurrences, especially among strains exhibiting distinct carbon consumption profiles. In this work, we have opted for the exclusive incorporation of negative interactions in Lotka-Volterra equations. We posit that the simpler and more justified assumption lies in considering competition for nutrients as the primary environmental-mediated interactions shaping bacterial populations.

Given the initial setup involves spatially separated bacterial clusters, adopting a metapopulation model proves advantageous. Metapopulation models are frameworks in ecology that conceptualize the dynamics of interconnected populations within a fragmented landscape. Coined by Richard Levins \cite{Levins}, the metapopulation concept views a population as a set of subpopulations occupying discrete patches of habitat, with occasional migration or dispersal occurring between these patches. The dynamics of each subpopulation are influenced by local factors like birth, death, and interactions, as well as the exchange of individuals among patches. Ngoc et al. offer an illustration of Lotka-Volterra formalism tailored for metapopulation in their work \cite{Ngoc}. Their model explores the dynamics of two species in competition for an implicit resource within a habitat divided into two patches.

In this work we have adapted the Lotka-Volterra formalism as follows:
\begin{equation}
\begin{split}
    \frac{dX_{i\alpha}(t)}{dt} = r_{i} X_{i\alpha}(t)\bigg(1 -  \frac{1}{K_i}\sum_{j\in\alpha} A_{ij} X_{j\alpha}(t)\bigg) \\
    + \mu\sum_{\beta}^{N_p}M_{\alpha\beta}\bigg(X_{i\beta}(t)-X_{i\alpha}(t)\bigg)
\end{split}
\end{equation}
where the Latin indices refer to species while the Greek ones to the different $N_p$ patches; thus, $X_{i\alpha}$ represents the population of species $i$ in patch $\alpha$. The intrinsic growth rate and the carrying capacity of species $i$ are indicated as $r_i$ and $K_i$ respectively. The interaction coefficient $A_{ij}$ depends only on the species types $i$ and $j$ and the sum in the first term runs over all the species $j$ placed in patch $\alpha$, i.e. species interact only if in the same patch. Moreover, in this formulation the entries of $\mathbf{A}$ are all positive in order to reproduce only competitive interactions. The final term, modulated by the coefficient $\mu$ incorporates diffusion when, for example, a species grows enough to spread to neighboring patches. The matrix $\mathbf{M}$ represents the network connecting the patches, obtained from the last random geometric graph provided by the simulation. In particular, $M_{\alpha\beta} = m_{\alpha\beta}/n_{\alpha}n_{\beta}$ with $m_{\alpha\beta}$ number of links between the communities and $n_\alpha$, $n_\beta$ number of nodes in $\alpha$ and $\beta$ respectively. In summary, the weights $M_{\alpha\beta}$ quantify the spatial proximity between different patches. 

\subsection{On the stability of Lotka-Volterra equations}

As $\mu$ approaches 0, the equilibrium solutions within each patch become well-established and are solely dependent on the interaction pattern and carrying capacities, given by $X_{i\alpha}^* = \sum_{j\in\alpha}(\mathbf{A}^{-1})_{ij}K_{j}$. It is recognized that for global stability of the feasible fixed point ($X_{i\alpha}^* > 0 \quad\forall\, i,\alpha$), the matrix $\mathbf{A}$ must be negative definite. In other words, $\mathbf{A} + \mathbf{A}^T$ should possess exclusively negative eigenvalues, as thoroughly explained by Grilli et al. in their study \cite{Grilli2}. In his groundbreaking research \cite{May}, Robert May demonstrated that large ecological networks exhibit a notably low probability of stability. Specifically, when matrix entries are sampled from a random distribution with a mean of zero and a mean square value of $\alpha$, the system is almost certain to be unstable if $\alpha > 1/\sqrt{N}$. Building upon May's findings, Allesina and Tang \cite{Allesina} extended the results differentiating between the various types of relationships, including predator-prey, competition, or mutualism. They provided analytical stability criteria for each scenario. The key takeaway from these insights is that in the case of competition, system stability is only assured when there is a predominant presence of very weak interactions. 

Mathematical analyses of stability typically focus on conditions near equilibrium points due to analytical challenges in dealing with nonlinear systems at a distance from equilibrium. For this reason, Holling \cite{Stability} suggested defining the behavior of ecological systems with two distinct properties: resilience and stability. Resilience pertains to the persistence of relationships within a system and measures its ability to absorb changes in state variables, driving variables, and parameters while still persisting. Stability, on the other hand, refers to a system's capacity to return to equilibrium after a temporary disturbance, with a more rapid and less fluctuating return indicating greater stability. The well-known \emph{mutual invasion criterion}, often associated with the concept of resilience, serves as a notable benchmark. For stable coexistence, it demands that each species within a community demonstrates positive population growth rates when invading a pre-existing community of competitors from low density \cite{invasioncriterion}. This is exactly the indicator used by Chang et al. to assess stable coexistence in their experimental communities of microbial species. Take a moment to focus on a scenario involving only two species within the system. In the context of competitive Lotka-Volterra equations, it is established that the invasion criterion holds true when $A_{12} < K_1/K_2$ and $A_{21} < K_2/K_1$ \cite{Macarthur}. Consequently, note that weak interactions with zero mean would almost always satisfy the invasion criterion and, consequently, lead to stable coexistence. However, Chang et al. \cite{doi:10.1126/science.adg0727} observed in their experiments that only a relatively small fraction (about 30\%) of possible pairs of species complies with the invasion criterion. When contemplating the parameters $A_{ij}$ as `universal' coefficients for pairwise interactions, i.e. only depending on species' types, it appears that these experimental results are scarcely consistent with the conditions necessary for mathematical asymptotic stability. 
The work by Abramson and Zanette \cite{Abramson}, provides us a workaround. They randomly selected interaction coefficients for a system comprising $N$ Lotka-Volterra species from a uniform distribution centered around one. Remarkably, the resulting phase space exhibited a multitude of fixed points, with a majority featuring both positive and negative eigenvalues—indicating instability. Consequently, the system traverses various unstable equilibria, leading to instances where the population of certain species undergoes pseudo-extinctions, reaching very low concentrations before rebounding. The noteworthy finding in their study is the demonstration that introducing a lower bound, $X_0$, to the populations induces a shift in the stability of nearly all equilibria, transforming them into stable states. It is important to note that the imposition of a lower threshold on populations is entirely justified. Indeed, the population density of a species or genotype confined to a specific spatial volume $V$ cannot fall below $V^{-1}$ unless it disappears entirely. When describing densities, it becomes essential to establish a threshold below which the density effectively approaches zero.
All of these heuristic discussions provide compelling justifications for our parameter choices in the model, which will be detailed in the results section. Moreover, we recall that our theoretical setup incorporates the structural aspect of patches, a factor that significantly influences coexistence and stability requirements. The next section will provide further clarification on this point.

\subsection{Mesoscopic Interpretation of the Interactions Parameters}

Let's consider the dynamics for the entire population \(X_i = \sum_{\alpha}X_{i\alpha}\) of the species rather than focusing on the subpopulations in individual patches. If we express the population of species \(i\) in terms of fractions \(\phi\) ranging from 0 to 1 (i.e., \(X_{i\alpha}(t) = \phi_{i\alpha}(t) X_{i}(t)\)), we obtain the following equation:
\begin{equation}
\begin{split}
\frac{dX_i(t)}{dt} = r_iX_i(t)\bigg(1-\frac{1}{K_i}\sum_{\alpha, j}\phi_{i\alpha}(t)A_{ij}\phi_{j\alpha}(t)X_j(t) + \\
\frac{\mu}{r_i}\sum_{\alpha, \beta}M_{\alpha\beta}(\phi_{i\beta}(t) - \phi_{i\alpha}(t))\bigg).
\end{split}
\end{equation}

Being the metapopulation network undirected, i.e. $M_{\alpha\beta} = M_{\beta\alpha}$, the last term of equation above is zero. Indeed, the migration between patches does not affect the global population. To recover the standard Lotka-Volterra equations for entire populations, we define
\begin{equation}
     A^G_{ij}(t)\equiv A_{ij}\,S_{ij}(t) = A_{ij}\sum_{\alpha}^{N_p}\phi_{i\alpha}(t)\,\phi_{j\alpha}(t).
     \label{eqn:proximity}
\end{equation}
Here, \(S_{ij}\), ranging between 0 and 1, measures the \emph{proximity} between species $i$ and $j$, i.e. the extent of segregation between them, indicating how much they occupy the same patches. When they don't share any patches, \(S_{ij}\) is zero; when their entire populations are in the same patch, \(S_{ij} = 1\). This interpretation of the global interaction parameters \(A_{ij}^G\) combines the `true' strength of interaction $A_{ij}$ between two species with the mesoscale spatial distribution across patches. It's noteworthy that even two strongly interacting species with a high \(A_{ij}\) can have a limited global impact on each other if they are adequately segregated in space. Indeed, the proximity measure changes the solutions at equilibrium which now have to satisfy
\begin{equation}
    \sum_j(\mathbf{S}^*\odot\mathbf{A})_{ij}X_j^*=K_i.
\end{equation}

Let's see how the invasion criterion in the case of two species changes in this new fashion. Suppose that species 1 is rare, $X_1\sim0$, while species 2 is at equilbrium $X^*_2$:
\begin{equation}
\begin{split}
    \frac{dX_2^*}{dt} = r_2X_2^*\bigg(1-\frac{1}{K_2}\sum_{\alpha}^{N_p}(\phi_{2\alpha}^*)^2A_{22}\,X_2^*\bigg)\\ = 0 
\Leftrightarrow \quad X_2^* = \frac{K_2}{A_{22}\sum_{\alpha}(\phi^*_{2\alpha})^2}.
\end{split}
\end{equation}
The species 1 will be able to invade the system if $dX_1(t)/dt > 0$, i.e. if
\begin{equation}
A_{12} < \frac{K_1}{K_2}\,\frac{\sum_{\alpha}(\phi^*_{2\alpha})^2}{\sum_{\alpha}\phi_{1\alpha}\phi_{2\alpha}^*}\equiv \frac{K_1}{K_2}\,\Omega_{21},
\label{eqn:invasion}
\end{equation}
where we have considered $A_{ii} = 1$. To comply with the mutual invasion criterion the same must hold also for $A_{21}$, inverting the indices. However species 2, in absence of species 1, has grown equally in all the patches until equilibrium; this means that $\phi^*_{2\alpha}=1/N_p,\,\,\forall\alpha$. Considering also that $\sum_{\alpha}\phi_{1\alpha} = 1$, we obtain that the factor $\Omega$ is equal to one and we revert to the classical invasion criterion conditions. This elucidates how the mesoscale structure can at the same time relevantly affect the global interactions \(A_{ij}^G\), and so the steady state, for species in communities while maintaining the same conditions for pairwise stable coexistence in the sense of the invasion criterion.

\subsection{Macroecological Laws}

A recent contribution by J. Grilli \cite{Grilli} represents a significant stride in the macroecological exploration of microbial communities. Through the analysis of data from different biomes, the study delineates patterns of abundance variation, encapsulating three macroecological laws: i) the abundance fluctuations of any given species across samples adhere to a gamma distribution; ii) the variances of these distributions for distinct species are proportional to the square of their means, known as Taylor's law \cite{Taylor}; and iii) the mean abundances across species conform to a lognormal distribution. He also showed how a Stochastic Logistic Model (SLM), which enhances a logistic growth with a multiplicative stochastic term reproducing enviromental effects, can perfectly reproduce the phenomenology. In an even more recent work by Camacho-Mateu et al. \cite{camachomateu2023sparse} a generalized stochastic Lotka-Volterra model (SLVM) was introduced, incorporating pairwise interactions. This model was also able to uphold Grilli's three empirical laws. They considered only weak interactions in order to comply with the global stability of the feasible fixed point requirements. We will demonstrate that our framework can replicate Grilli's laws, even when accounting for stronger interactions and incorporating information about environmental stochasticity within our spatial simulation.

\section{Results}
We conducted multiple runs of our spatial simulation, considering $n=2000$ bacteria, i.e. nodes in the RGG, assigned to $N=100$ different species and exploring various values of the radius \(R\) and threshold \(N_{th}\). Remarkably, for each radius value, it is always possible to find a corresponding \(N_{th}\) such that the simulation reaches a form of equilibrium relatively quickly. This equilibrium comprises different spatially segregated clusters of bacteria, where each bacterium attains a very low velocity of motion and predominantly occupies a single patch.
\begin{figure*}
   \hspace*{-2cm} 
    \includegraphics[scale = 0.43]{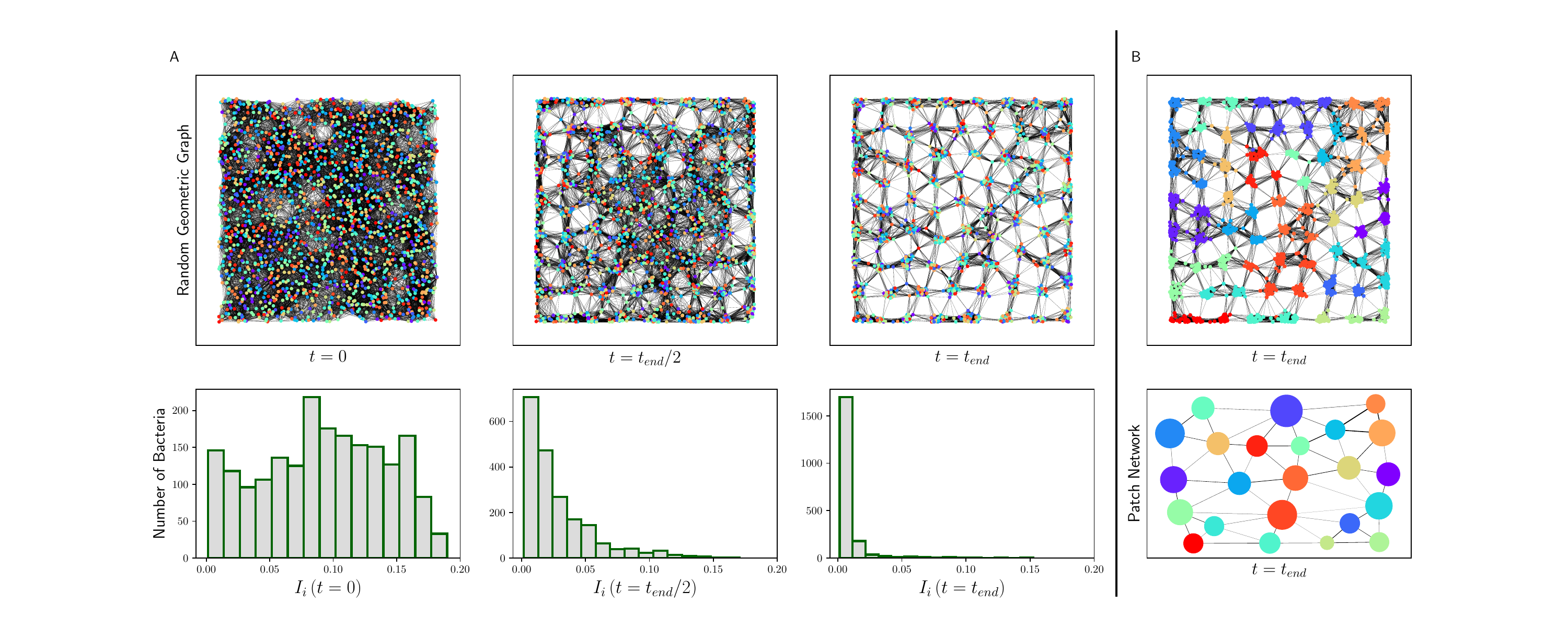}
    \caption{In this example the individual-based simulation was conducted in a two-dimensional square of dimensions \(2 \times 2\) with \(R = 0.2\), $\alpha = 7$ and \(N_{th} = 60\). We considered $n= 2000$ nodes (bacteria) randomly assigned to $N=100$ different species. The top row of panel A) presents the random geometric network representation of the system where each species is indicated with a different color, while the bottom row illustrates the intensity motion distribution as described by equation \ref{eqn:intensity}, both showed at three different time-steps ($t=0$, $t=t_{end}/2$ and at the end of the simulation $t_{end}$). In panel B), we present the outcome of the community detection algorithm. Initially, we depict the identified communities using distinct colors (top row). Subsequently, we construct a patch network wherein communities are represented by individual nodes of corresponding colors (bottom row). In this network the size of each node is proportional to the number of bacteria within the community and the edges are weighted based on the number of links connecting the communities.}
    \label{fig:spatial}
\end{figure*}
Figure \ref{fig:spatial} presents an example of simulation with $R = 0.2$, $\alpha=7$ and \(N_{th} = 60\), with snapshots of the system at different time steps. In panel A) The RGG is displayed alongside the associated intensity motion distribution. As the nodes cluster, the distribution shifts towards very low values. This phenomenon persists across all the higher values of $R$ investigated, extending up to 0.5. The straightforward outcome of increasing the distance in the RGG is a reduction in the number of patches and an increase in their dimensions. Consequently, for the subsequent analysis, we focus solely on the case where $R=0.2$. All the simulations are performed in $2\times2$ bidimensional square. It's important to note that the range of velocities considered is biologically plausible: bacteria can move at a wide range of speeds ranging from $1\,\,\mu m/s$ to $1000\,\, \mu m/s$ \cite{Berg2}; considering the square as $2 \times 2 \,\,mm^2$, this implies that bacteria at the beginning of our simulation travel on average 141 $\mu m$ in a few seconds of straight-line motion, with a maximum possible value of about 280 $\mu m$; then, at the end of simulation, they all reach velocities of very few $\mu m$ per second.

Upon reaching equilibrium, each simulation yields the initial distribution of various species across patches, representing the initial conditions in our metapopulation model for species growth. We numerically solved the system of differential equations using an explicit Runge-Kutta method of order 5. For simplicity, we opted for identical and unitary intrinsic growth rates and carrying capacities for all species ($K_i = 1,\,\, r_i = 1,\,\,\forall i$). In light of section III.A, we chose not to require asymptotic stability but to follow a direction similar to Abramson et al. \cite{Abramson}, therefore we randomly sampled the off-diagonal entries of the interaction matrix from a lognormal distribution with a mean ($\mu$) set to one, varying its variance ($\sigma^2$) from 0.05 to 0.5 ($A_{ij}\sim\text{Lognormal}(\mu, \sigma)$). The diagonal elements were uniformly set to one ($A_{ii} = 1$). Additionally, we incorporated a lower limit for the population, denoted as $X^0_{i\alpha}$, set at $10^{-3}$. Furthermore, we explored various values for the diffusion parameter ($\mu$), spanning from zero to one.
%\begin{figure*}
 %   \centering
  %  \includegraphics[scale=0.6]{Figure3.pdf}
   % \caption{Graphical description of the simulation process, which aims to replicate the experimental procedure outlined by Chang et al. Panel A) Displays the Random Geometric Graph (RGG) at the conclusion of the spatial simulation. Panel B) Illustrates the different communities identified by the Louvain algorithm, distinguished by various colors. Panel C) Shows the corresponding patch network, where the size of each patch is proportional to the number of bacteria it contains, and the links are weighted according to the number of connections between elements within the patches. Panel D) Depicts the temporal dynamics of relative abundances for each species, obtained from the numerical solution of the metapopulation Lotka-Volterra equations. Following each simulation, we identify the species that survive at equilibrium and subsequently consider all potential pairs of these species, re-executing the entire model, including spatial simulation (panel E) and growth (panel F), using only those pairs. Mirroring the experimental setup, we explore three distinct initial distribution proportions for the pairs (50\%-50\%, 95\%-5\%, 5\%-95\%) to determine if they can coexist under each scenario.}
 %   \label{fig:schema}
%\end{figure*}
Confirming the findings of Abramson et al. \cite{Abramson}, in this setup, we observed that the numerical simulations display equilibrium. We considered the system to exhibit equilibrium dynamics if the coefficient of variation for the populations of the species was below 1\% in the last 500 time-steps. This criterion was met in the majority of cases, while in some instances, species' populations exhibited oscillatory behavior. We focused solely on the former cases for our analysis. To assess the stability of these equilibrium configurations, we employ the mutual invasion criterion. This involves systematically reducing the abundance of each surviving species at equilibrium one by one and observing whether they exhibit a positive growth rate, ultimately returning to values close to their previous abundance levels. We observed that approximately 90\% of the surviving species consistently demonstrate regrowth potential in all the cases investigated, guaranteeing a form of stability for our simulated systems. Those species that became extinct were primarily those with already minimal abundance levels at equilibrium. 

The idea is to organize the simulations to replicate the procedure in \cite{doi:10.1126/science.adg0727}: the spatial simulation ends up with the RGG at equilibrium, on which we employed the modularity optimization Louvain algorithm which successfully identifies distinct communities, i.e. the different patches of the metapopulation network (panel B of figure \ref{fig:spatial}). Afterwards, we use the metapopulation Lotka-Volterra formalism to simulate species growth and to identify the species that reach equilibrium. Subsequently, we proceed to re-execute the entire model again, including both spatial simulation and growth, considering all the possible pairs of survivors. Consistent with the experimental design, we investigate three different initial distribution proportions for the pairs (50\%-50\%, 95\%-5\%, 5\%-95\%) to assess their potential for coexistence under each scenario in the absence of the other species.

\begin{figure*}
    \centering
    \includegraphics[scale = 0.7]{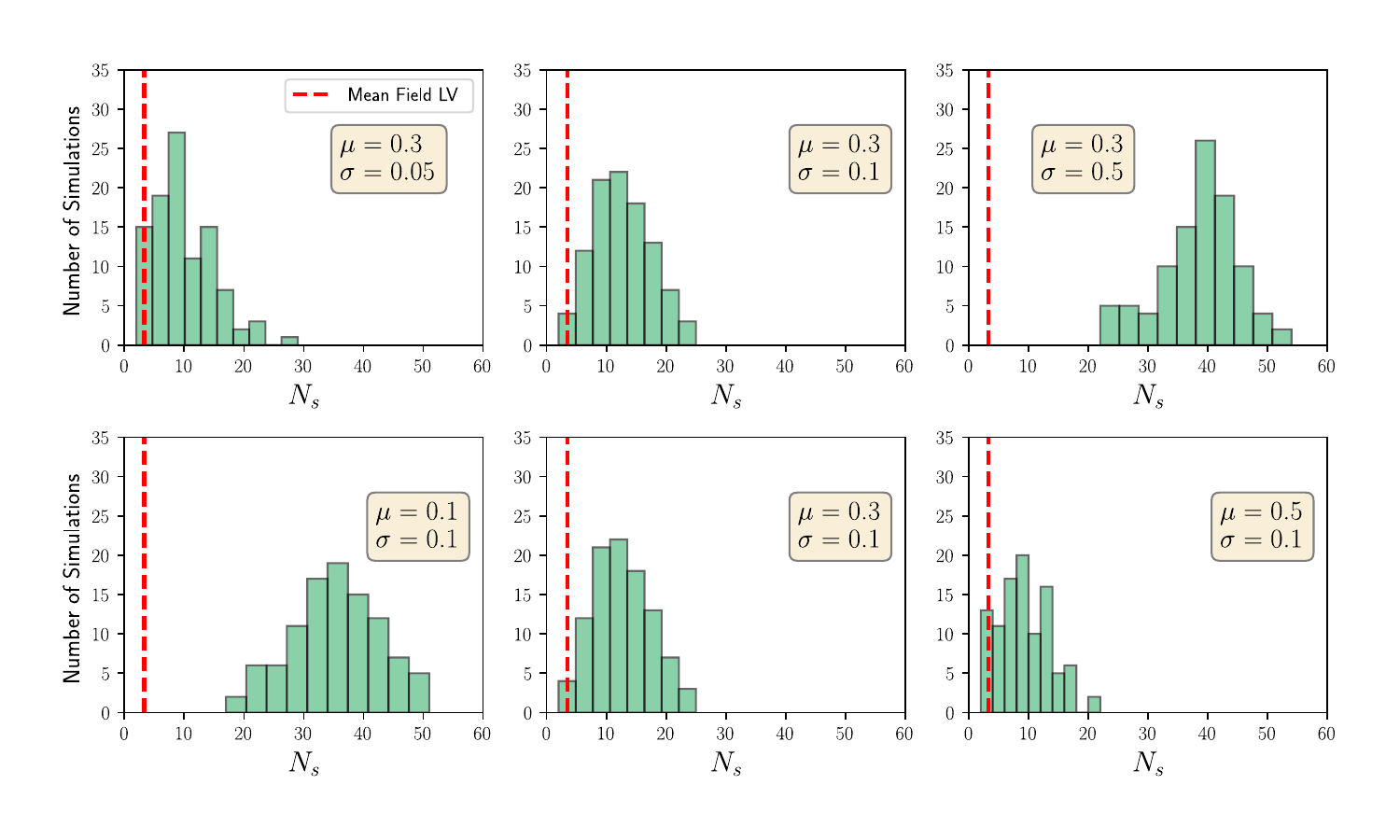}
    \caption{Distributions of the number of survivors ($N_S$) obtained after 100 runs of the spatial simulation and the metapopulation model for three different values of the interactions standard deviation $\sigma = 0.05,\,0.1,\,0.5$, while keeping $\mu = 0.3$ in the top row and for $\mu = 0.1,\,0.5,\,0,5$ while keeping $\sigma  = 0.1$ in the bottom one. The vertical dashed red line indicates the mean number of surviving species over 100 runs of the classical Mean-Field Lotka-Volterra model (Mean Field LV). For all the simulations $n = 2000$, $N = 100$,\,$R = 0.2$, $N_{th} = 60$, $\alpha=7$, $K_i = 1$ and $r_i = 1\,\forall i$, $X^0_{i\alpha} = 10^{-3}\,\forall \,(i,\,\alpha)$.}
    \label{fig:histo}
\end{figure*}

The initial series of simulations consisted of 100 runs, considering different combinations of the interaction standard deviation ($\sigma$) along with the diffusion parameter ($\mu$). We examined three values of $\sigma$ (0.05, 0.1, and 0.5) and three values of $\mu$ (0.1, 0.3, and 0.5). Figure \ref{fig:histo} displays the outcomes for the number of survivors ($N_S$) while varying $\sigma$ and keeping $\mu$ constant (top row), and vice versa (bottom row). A vertical dashed red line indicates the mean number of surviving species over 100 runs of the classical Mean-Field Lotka-Volterra without patch structure. Notably, across all cases, the patch structure consistently facilitates a significantly higher number of surviving species compared to the mean field scenario, indicative of enhanced coexistence among species. The distribution shifts towards higher values for the number of survivors with higher $\sigma$ and lower diffusion rates $\mu$. This latter trend can be explained straightforwardly: as $\mu$ increases, species can explore the entire patch network more rapidly, converging towards the classical mean-field approach. Conversely, when variance increases, the distribution exhibits heavier tails, and the correlation between higher variances and a greater number of survivors suggests that increased heterogeneity in the interaction patterns among species, and therefore in the composition of the patches, can promote coexistence. We will revisit the effect of the heterogeneity later in our analysis.

\begin{figure}
    \centering
    \includegraphics[scale = 0.6]{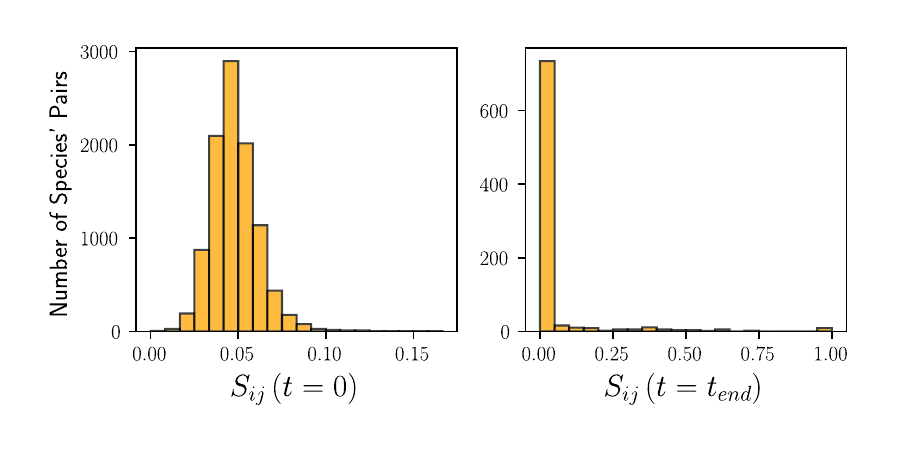}
    \caption{Distribution of the proximity measure $S_{ij}$, as defined in equation \ref{eqn:proximity}, for each possible pair of species, before and after species' growth and over 100 runs of the complete model. Given that  $S_{ij}$ is not defined when one or both species are extinct, in the final state we considered all possible couples of only survived species. For the simulation here $n = 2000$, $N = 100$,\,$R = 0.2$, $N_{th} = 60$, $\alpha=7$, $\sigma = 0.1$, $\mu = 0.3$, $K_i = 1$ and $r_i = 1\,\forall i$, $X^0_{i\alpha} = 10^{-3}\,\forall \,(i,\,\alpha)$. }
    \label{fig:segreg}
\end{figure}

Further insights can be obtained looking to the fractions $\phi_i$ before and after executing the model. Specifically, refer to figure \ref{fig:segreg} to observe the `proximity measure' $S_{ij}$, as defined in equation \ref{eqn:proximity}, for each pair of species both before and after allowing them to grow and migrate. In the initial state, $S_{ij}$ exhibits a bell-shaped distribution, indicating that the spatial simulation results in species that are, on average, evenly segregated among the patches. Following the model execution, the surviving species demonstrate a distribution more concentrated around zero and with a long tail, suggesting that survivors are more likely to be highly segregated with a small minority of them being able to coexist in the same patches. As described by equation \ref{eqn:proximity}, this phenomenon facilitates the coexistence of species, even when their net interaction $A_{ij}$ is high. 

\begin{figure}
    \centering
    \includegraphics[scale=0.6]{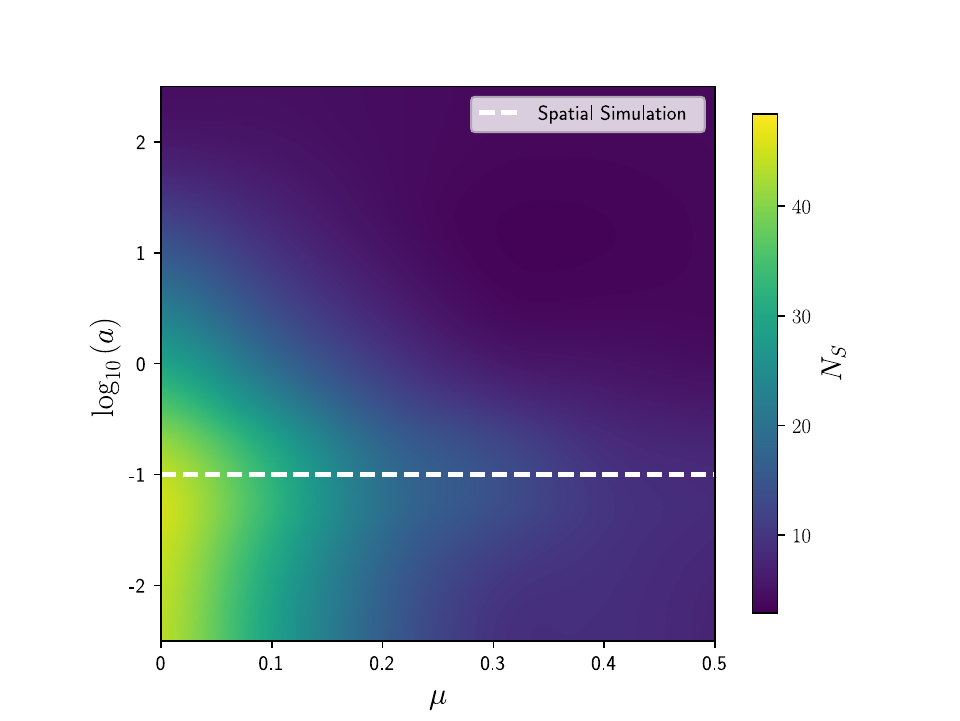}
    \caption{The heatmap depicts the number of survivors ($N_S$) at the equilibrium varying the initial distribution of the species, through the concentration parameter $a$ in a Dirichlet distribution (y-axis in log scale), and the diffusion rate $\mu$ (x-axis). Brighter colors indicate higher values for the number of survivors. The white horizontal dashed line indicates the value of $a$ which best fits with the initial distribution of the species provided by our spatial simulation. For each pair of values $(a, \mu)$ we performed 100 runs of the model with $\sigma = 0.1$, $K_i = 1$ and $r_i = 1\,\forall i$, $X^0_{i\alpha} = 10^{-3}\,\forall \,(i,\,\alpha)$. To ensure the figure appears continuous, we used quadratic interpolation.}
    \label{fig:heatmap}
\end{figure}

Successively, we aimed to assess the influence of the initial distribution of species across the patches on the number of coexisting species at equilibrium. To achieve this, rather than conducting our spatial simulation, we sampled the initial populations from a Dirichlet distribution to ensure equal abundances for all species ($\sum_{\alpha=1}^{N_p}X_{i\alpha}(0)=1$), but with varied distributions among the patches
\begin{equation}
    \vec{X_{i}}(0) = (X_{i1}(0), X_{i2}(0),\dots, X_{iN_p}),
\end{equation}
\begin{equation}
    p(\vec{X_i}(0)) = \frac{1}{\mathcal{N}}\prod_{\alpha = 1}^{N_p}X_{i\alpha}^{a_\alpha-1}(0),
\end{equation}
with $\mathcal{N}$ normalization factor (multivariate beta function) and $\vec{a}=(a_1,\dots,a_{N_p})$ concentration parameters for the different patches. We chose to consider equal $a_{\alpha} = a,\,\,\forall\alpha$, in order to give the same importance to all the patches. Lowering the global parameter $a$ introduces greater random variability in the species' concentration among the patches and, therefore, higher heterogeneity in the patch composition. We conducted multiple model runs, varying the value of $a$ from 100 to 0.01 while simultaneously adjusting the diffusion rate to generate the heatmap depicted in figure \ref{fig:heatmap}.
The figure highlights that lower values of the diffusion rate and greater variability between patch compositions (lower $a$) yield the best performance in terms of the number of surviving species. This further suggests that heterogeneity promotes coexistence among multiple species. Upon attempting to describe the initial distribution of species yielded by our spatial simulation with the Dirichlet distribution, we find that the value of $a$ which best fits it is approximately 0.1 (indicated by the dashed horizontal line in Figure \ref{fig:heatmap}). This indicates that the spatial simulation effectively reproduces highly heterogeneous patches.

\begin{figure}
    \centering
    \includegraphics[scale = 0.7]{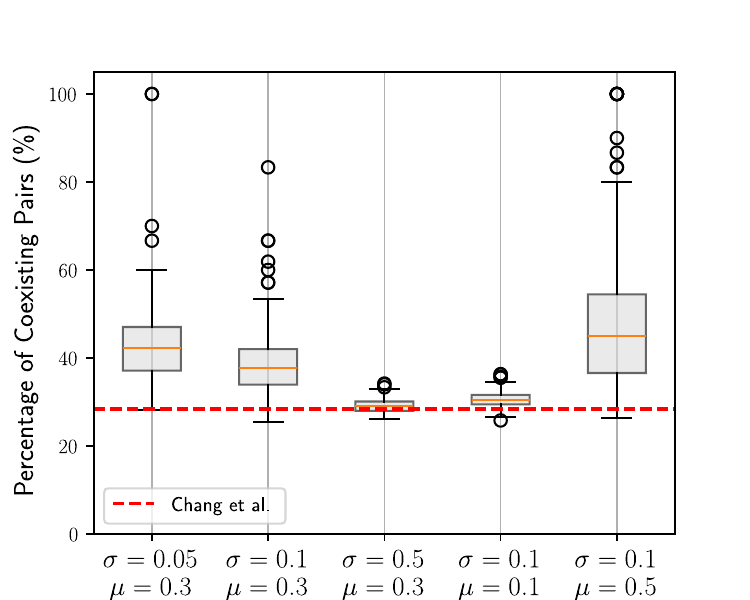}
    \caption{
    Each box represents the distribution for the percentage of coexisting species' pairs (picked from the set of survivors in the multi-species simulation) over 100 runs for different combinations of $\mu$ and $\sigma$, as depicted along the x-axis. The box extends from the first quartile (Q1) to the third quartile (Q3) of the data, with a line at the median. The whiskers extend from the box to the farthest data point lying within 1.5x the inter-quartile range (IQR) from the box. Flier points indicate outliers. The red horizontal dashed line indicates the experimental value for the percentage of coexisting pairs found in \cite{doi:10.1126/science.adg0727}. For the multi-species part the chosen parameters are $n = 2000$, $N = 100$,\,$R = 0.2$, $N_{th} = 60$, $K_i = 1$ and $r_i = 1\,\forall i$, $X^0_{i\alpha} = 10^{-3}\,\forall \,(i,\,\alpha)$. The same for the two-species part except for $N=2$.}
    \label{fig:boxplot}
\end{figure}

Interesting enough, upon re-executing the model for all possible pairs of surviving species after all simulations, we discovered that only a minority of pairs can coexist in isolation, mirroring the experimental findings and replicating the emergent behavior. The boxplot in Figure \ref{fig:boxplot} illustrates the distributions of the percentage of species' pairs coexisting in isolation for the same combinations of $\mu$ and $\sigma$ as previously examined. A horizontal dashed line represents the value reported by \cite{doi:10.1126/science.adg0727}, which stands at 28.4\%. Specifically, cases with the highest $\sigma$ and the lowest $\mu$ demonstrate a significantly high probability of replicating the experimental value. For the remaining cases, although the average probabilities exceed those observed experimentally, again only a minority of pairs exhibit coexistence.
\begin{figure}
    \centering
    \includegraphics[scale = 0.425]{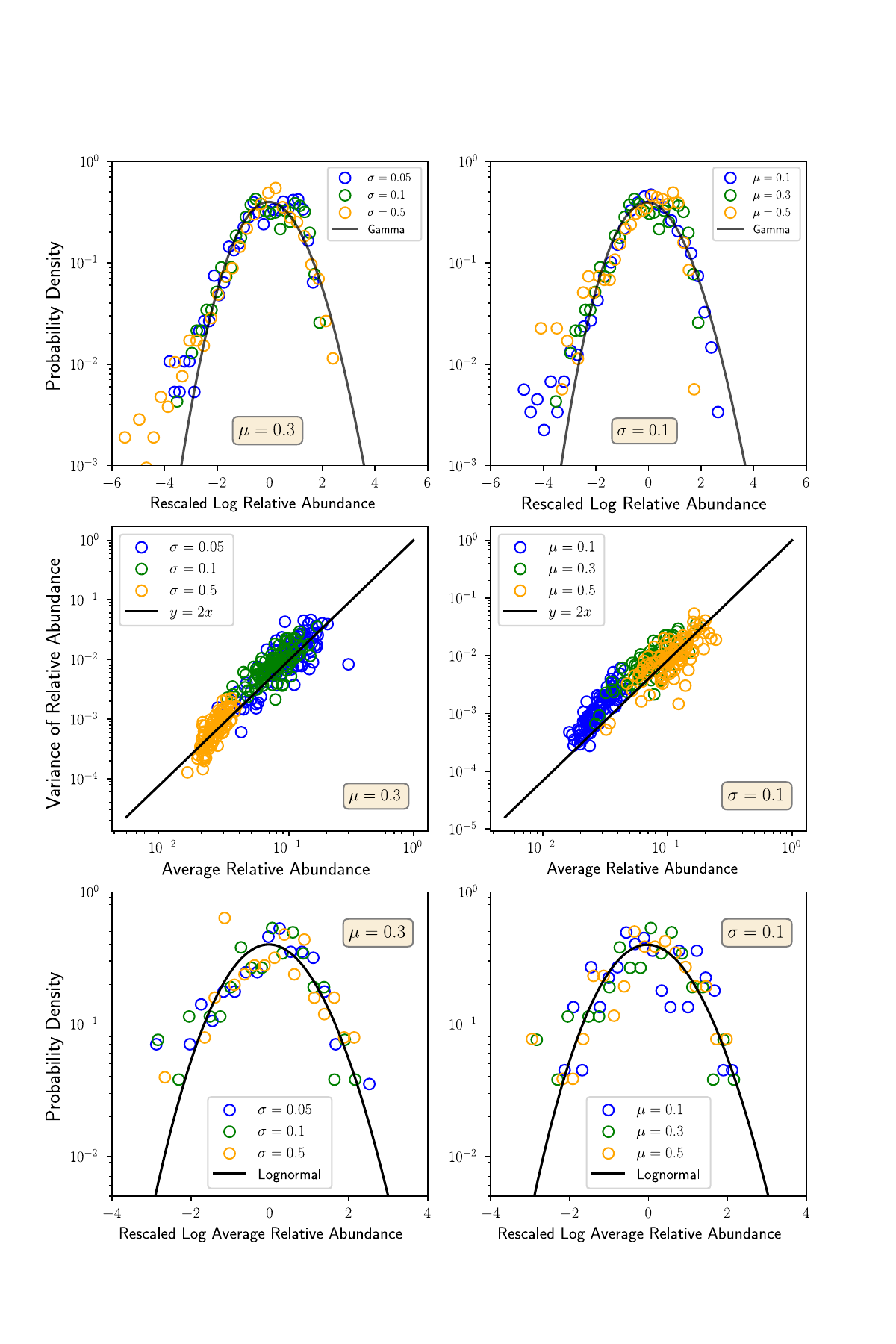}
    \caption{
    The simulations were conducted with $\sigma=0.1$ and varying $\mu$ across 0.1, 0.3, and 0.5 for the plots in the right column. Similarly, the simulations were repeated with $\mu=0.3$ while adjusting $\sigma$ to 0.05, 0.1, and 0.5 in the left column. The different colors of the circles represent the three values of $\mu$ in the right column and of $\sigma$ in the left column. The remaining parameters in the model were set as in Figure \ref{fig:histo}. The top row presents the Abundance Fluctuation Distribution, depicting the probability density for the rescaled log relative abundance of the species across 100 model runs. Here, the black line represents a gamma distribution. In the second row, we plotted the variance of the relative abundance as a function of the corresponding means across 100 simulations. The black line corresponds to Taylor's law with an exponent of two. The third row displays the distribution for the rescaled log average relative abundance, again across 100 different simulations, compared with a lognormal distribution (black line). The analysis yielded mean $R^2$ values of 0.97, 0.5, and 0.86, respectively, for the three laws.}
    \label{fig:grilli}
\end{figure}
In fact, in the absence of multiple species, spatial self-organization in this scenario leads to a random process distributing both species equally across all patches with high probability. This implies an unaffected interaction pattern, rendering the spatial distribution incapable of promoting coexistence, in contrast to what happen in the case of multiple species, where the spatial simulation naturally generates more heterogeneous patches. When there is a significant disparity in the initial concentration of the two species, as previously mentioned, the patch structure does not impact the invasion criterion when one species invades the other. Consequently, the mesoscale structure, while potentially influencing coexistence in multi-species configurations, does not change the criteria for coexistence in the case of two species. This elucidates the findings illustrated in Figure \ref{fig:boxplot}.

As mentioned earlier, a robust testing ground for a model aspiring to describe the microbiome involves verifying its ability to replicate the three macroecological laws outlined in section III.C. To achieve this, we conducted 100 model runs, again for different combinations of the interactions variance and the diffusion rate and then we checked for the abundance distribution of species across samples (Abundance Fluctuation Distribution, AFD), the relationship between the variance and mean of species abundances, and the distribution of mean abundances across species (Mean Abundance Distribution, MAD). In the left column of figure \ref{fig:grilli} the three laws are displayed, considering $\sigma$ fixed to 0.1 and for three different values of $\mu = 0.1,\,0.3,\,0.5$. In the right column we did the same but fixing $\mu$ to 0.3 and for three values of $\sigma = 0.05,\,0.1,\,0.5$. The three laws are depicted along the rows. All these analyses demonstrated good agreement with the patterns identified by J. Grilli, resulting in mean $R^2$ values of 0.97, 0.5, and 0.86, respectively, for the three laws. The fits are performed considering a number of bins of the distributions for the first and the third law according to the Freedman-Diaconis rule \cite{freedman}.

The second law exhibits less alignment with theoretical expectations. Nevertheless, recent findings \cite{JoseCuesta} strongly indicate that the second law cannot be regarded as a precise relationship between mean and variance. Instead, the exponent appears to be sampled from a distribution rather than remaining a constant, leading to increased dispersion in data points, as observed in our case. This hypothesis finds support in both empirical observations and the predictions of an approximate theory.

\section{Discussion}

In this study, we have developed a theoretical framework that combines an individual-based spatial simulation of bacterial motion with a metapopulation Lotka-Volterra model to describe abundance dynamics. The current model has demonstrated its capability to replicate certain experimental phenomena, such as the intriguing emergence of coexistence observed by Chang et al. Specifically, the model facilitates coexistence in multi-species systems by fostering the spatial rearrangement of bacteria into distinct clusters. This spatial segregation is a consequence of a high escape response triggered when the local environment becomes densely populated and competitive. The self-organization into different patches consistently leads to a greater number of bacteria surviving, as spatial separation limits interactions among them. However, this process proves to be less efficient in the case of a two-species system. When two species have a strong interaction, the only viable way to coexist is complete segregation within the system. If bacteria are uniformly distributed, there is a high probability that both species will end up in each patch, therefore not changing the pattern of competition. In summary, our spatial simulation demonstrates the capability to generate heterogeneous patches in scenarios involving multiple species, whereas this heterogeneity is not observed in cases with only two isolated species. This discrepancy results in the promotion of coexistence in the former case, while not in the latter.

One could posit that the described phenomenon occurs because bacteria, when assessing the density of surrounding competitors, are unable to discriminate between different species. This leads to the inclusion of cells from the same species as perceived competitors. However, there is no inherent reason to assume any form of cooperation between bacteria of the same species. As mentioned earlier, cooperation, although possible, tends to be infrequent both among bacteria of the same species and across different species \cite{Hibbing, Palmer, Foster}. This rationale underlies our decision to exclusively consider negative interactions in the formulation of the growth model.

Similarly, it can be argued that the absence of uneven nutrient distribution in our model may impose limitations on its applicability. Nevertheless, when considering experimental communities, they are typically cultured in agar plates or batch cultures where nutrients are generally thoroughly mixed, resulting in a uniform distribution throughout the medium. Given this context, we assume that there is no substantial nutrient concentration gradient influencing bacterial motion. Thus, we posit that the predominant driving force for movement is the avoidance of competition and we showed  how even in isotropic environments with no particular nutrients distribution, is still possible to have a spatial structure with meta-communities, affecting the coexistence among species. A consumer-resource model would necessitate making various assumptions about nutrient abundances, types, temporal dynamics, spatial distribution, and more. The Lotka-Volterra formalism allows the modeling of resource competition without delving into specific resource details. 
%In Chang et al.'s experiments, nutrients start off limited, leading to bacterial competition. However, the nutrients undergo repeated replenishment throughout multiple cycles. If we consider the experiment duration significantly exceeds the duration of a single cycle, it is akin to assuming an overall stable nutrient concentration. Consequently, even if a different model, like MacArthur's Consumer-Resource model incorporating nutrient dynamics, were employed, assuming globally constant nutrient concentrations over time would essentially reduce it to a formalism equivalent to Lotka-Volterra dynamics.

Bacteria rank among the fastest reproducers globally, doubling at the scale of minutes \cite{Allen}. However, their motion is also remarkably rapid, with swimming speeds exceeding 100 body lengths per second \cite{Mitchell, LUCHSINGER19992377}. This substantiates our assumption of effectively separating the temporal scales between self-organization in space and growth.

While numerous instances of empirical evidence support bacterial self-organization in space, our model currently lacks experimental validation. Nevertheless, we have shown that our theoretical framework aligns with the three macroecological laws identified by J. Grilli, providing a form of empirical confirmation. Efforts are underway to experimentally validate our assumptions and to scale the model for increased biological plausibility. Future endeavors will involve a significantly higher number of bacteria situated in a three-dimensional space with varied geometry, a task currently beyond reach due to the computational power at our disposal. 

In conclusion, we believe that our approach is innovative as it validates the utilization of population-based models integrating meso-scale structures. While a fully microscopic approach  pose technical challenges, a mean-field approach would oversimplify, and existing works incorporating meso-structures often do not provide generative mechanisms for them. We introduced a microscopic process that potentially rationalizes the adoption of patch configurations. Simultaneously, we illustrated how minimal adaptability levels can lead to the formation of isolated bacterial clusters, maintaining a high degree of randomness that mirrors the intricate stimuli influencing bacterial motion. We believe that this work has the potential to pave the way for a new research direction, emphasizing the importance of considering the delicate spatial equilibrium between species within a microbial community as a pivotal element to be incorporated into theoretical models and investigations.

%In summary, our objective with this work is to convey a fundamental message — that space distribution is a crucial factor. Particularly in ecological theoretical models, the common approach involves a mean-field approximation, assuming that each species interacts with all others in the system. While a completely microscopic approach may pose challenges, our study demonstrates that incorporating a mesoscale structure, such as distinct communities or patches in space, can profoundly influence the dynamics and coexistence patterns.

%HIGHER ORDER INTERACTIONS? CAN THE SIMULATION IN SPACE BE SEEN AS AN HIGHER ORDER EFFECT? The presence of other species affects the enviroment
%TALK ABOUT MOTILITY AND ABOUT THE FACT THAT NOT ALL THE BACTERIA CAN MOVE BUT THOSE IN THE EXPERIMENTS YES

\section*{Author contributions statement}
A.A. and M.M. designed the study; M.M. performed simulations, mathematical calculations and data analysis; A.A. supervised the results; A.A. and M.M. wrote and approved the manuscript. 
\section*{Acknowledgements}
This project has received funding from the European Union's Horizon 2020 research and innovation programme under the Marie Skłodowska-Curie grant agreement No. 945413 and from the Universitat Rovira i Virgili (URV).
\section*{Competing interests}
The authors declare no conflict of interest.

\bibliographystyle{unsrt}
\bibliography{bibliography.bib}
%\end{multicols}

\end{document}